\documentclass[conference]{IEEEtran}
\usepackage{algorithm,algorithmic}
\usepackage{amssymb}
\usepackage{amsmath}
\usepackage{dsfont}
\usepackage{cite}
\usepackage{url}
\usepackage{xcolor}
\usepackage{cite,graphicx,amsmath,amssymb}
\usepackage{subfigure}
\usepackage{fancyhdr}
\usepackage{mdwmath}
\usepackage{mdwtab}
\usepackage{amsthm}
\usepackage{multirow}
\usepackage{flafter}
\usepackage[
top    = 1.70cm,
bottom = 1.05in,
left   = 0.63 in,
right  = 0.63 in]{geometry}

\newtheorem{theorem}{Theorem}

\newtheorem{lemma}{Lemma}

\newtheorem{corollary}{Corollary}

\makeatletter
\def\ScaleIfNeeded{%
\ifdim\Gin@nat@width>\linewidth \linewidth \else \Gin@nat@width
\fi } \makeatother

\IEEEoverridecommandlockouts
\begin{document}

\title{Joint User Activity and Data Detection in Grant-Free NOMA using Generative Neural Networks}

\author{
\IEEEauthorblockN{ Yixuan~Zou,~\IEEEmembership{Student Member,~IEEE,} Zhijin~Qin,~\IEEEmembership{Member,~IEEE,} Yuanwei~Liu,~\IEEEmembership{Senior Member,~IEEE,}\\} \IEEEauthorblockA{
Queen Mary University of London, London, UK\\
\{yixuan.zou, z.qin, yuanwei.liu\}@qmul.ac.uk
} 
}




\maketitle
\begin{abstract}
Grant-free non-orthogonal multiple access (NOMA) is considered as one of the supporting technology for massive connectivity for future networks. 
In the grant-free NOMA systems with a massive number of users, user activity detection is of great importance. Existing multi-user detection (MUD) techniques rely on complicated update steps which may cause latency in signal detection. In this paper, we propose a generative neural network-based MUD (GenMUD) framework to utilize low-complexity neural networks, which are trained to reconstruct signals in a small fixed number of steps. 
By exploiting the uncorrelated user behaviours, we design a network architecture to achieve higher recovery accuracy with a low computational cost. Experimental results show significant performance gains in detection accuracy compared to conventional solutions under different channel conditions and user sparsity levels. We also provide a sparsity estimator through extensive experiments. Simulation results of the sparsity estimator showed high estimation accuracy, strong robustness to channel variations and neglectable impact on support detection accuracy.
\end{abstract}

\section{Introduction}
The emerging Internet-of-Things (IoT) technologies accelerate the urge for reliable high-capacity wireless communication systems in supporting the massive connections among IoT devices and users. It brings challenges to the traditional grant-based orthogonal multiple access (OMA) schemes due to the limiting capacity of the conventional cellular systems and the high signalling overhead caused by the complicated and time-consuming scheduling. To overcome the aforementioned challenges, grant-free NOMA techniques have been proposed, which allow multiple users to share common resources non-orthogonally using controllable interference without complicated grant mechanism~\cite{NOMA_for_5G_2015}. 

A typical grant-free NOMA system can support more users than the number of orthogonal resources~\cite{NOMA_for_5G_2015} and data packets can be transmitted in the next available time slot without waiting for a grant. Since the BS has no prior knowledge of user activity, MUD should be performed at the BS in addition to signal detection. 
Exploiting the sporadic user activity, compressed sensing (CS) has been considered as a standard framework for solving the MUD problems
~\cite{sparse_representation_Qin2018}. 
In practice, users transmit information in a continuous manner and stay in the communication system for several time slots. Hence, as a mean to further improve the reconstruction performance, many recent works utilize the temporal correlations in user activity and adopt the \textit{frame-wise joint sparsity model} in their MUD algorithms~\cite{block_sparsity_MUD_NOMA_Du2018}~\cite{sparsity_blind_greedy_Yu2019}. In particular,~\cite{block_sparsity_MUD_NOMA_Du2018} proposed a threshold
aided block sparsity adaptive subspace pursuit algorithm to improve recovery accuracy and~\cite{sparsity_blind_greedy_Yu2019} proposed a sparsity-blind greedy algorithm to accommodate practical scenarios. However, both approaches rely on complicated update steps where the total number of iterations per recovery is unknown in either method. It raises the concern of slow recovery speed, which is critical to the latency-sensitive IoT applications.

Recently, a deep learning-based CS-MUD solution has been proposed, which achieves faster recovery speed and higher detection accuracy compared to conventional iterative algorithms~\cite{fast_CS_MUD_YannaBai2019}. This approach, however, requires multiple trained networks to be stored at the BS to accommodate different system overloading ratios, since the received signals are used as the network inputs.~\cite{DNN_MUD_NOMA_Kim2020} proposed a deep learning-based active user detection which detects sparse supports without prior knowledge on sparsity. Similar to~\cite{fast_CS_MUD_YannaBai2019}, this approach also uses the received signals as network inputs, resulting in the same requirements of different models for a different number of orthogonal resources.  

Compared to the typical neural network designs, 
generative networks, which could deal with arbitrary input dimensions, serve as a more flexible solution with fewer restrictions on applicable scenarios. 
To address the slow recovery speed of generative models in CS problems,~\cite{DCS_DeepMind} applied model-agnostic meta-learning (MAML) in the training algorithm so that the 
target signals can be recovered in as few as three iterations instead of hundreds. The fast and accurate signal recovery ability motivates the use of generative neural networks in solving MUD problems. However, standard generative network architectures~\cite{DCGAN_Radford2016} are not suitable for MUD. In typical communication system setups, users are assumed to behave independently and transmit signals based on random choices which make it difficult for generative networks to learn compared to image datasets which exhibit strong spatial correlations. Fortunately, the learning capability of generative networks can be improved by larger input dimensions~\cite{size_noise_tradeoff_GAN_Bailey2018}. However, the sizes of standard generative networks scale badly with input dimensions since fully-connected layers are commonly used as the first layer. 
Moreover, large kernel sizes are generally used in standard generative networks to capture the spatial features, whereas, in communication data, there are limited spatial correlations hence small kernel sizes are preferred.  

In practice, it is usually difficult to gain knowledge of user sparsity prior to MUD. Various efforts have been made on sparsity-blind CS approaches and a common solution is to estimate the signal sparsity adaptively in the recovery algorithm~\cite{block_sparsity_MUD_NOMA_Du2018}~\cite{sparsity_blind_greedy_Yu2019}. However, these sparsity approximation methods are each restricted to a certain type of recovery algorithms
such as subspace pursuit-based approaches~\cite{block_sparsity_MUD_NOMA_Du2018} 
and orthogonal matching pursuit (OMP)-based approaches~\cite{sparsity_blind_greedy_Yu2019}. Therefore, it is difficult to adopt the existing sparsity estimation techniques in new CS recovery algorithms, especially for the non-iterative deep learning-based approaches. 

In this paper, we propose a generative neural network-based MUD (GenMUD) framework based on the training algorithm in~\cite{DCS_DeepMind}. The main contributions are as follows:
\begin{enumerate}
    \item We propose to use generative networks in solving MUD problems to cope with different overloading factors with a single trained model.
    \item We design a low-complexity generative network architecture using 1x1 convolutional layers which significantly reduce the computational cost as input dimensions are increased to achieve higher recovery accuracy. 
    \item We obtain a closed-form low-complexity sparsity estimator which can be applied to any MUD algorithm to realize sparsity-blind MUD without complicated and iterative sparsity approximations.
    \item We show that the proposed approach outperforms conventional methods in terms of detection accuracy, and the proposed sparsity estimator achieves high accuracy under various channel conditions while making little impact on support detection accuracy.
\end{enumerate}


\section{System Model}\label{sec:system model}

\vspace{-0.5em}
\subsection{System Setup}
We consider the uplink NOMA system with K users and one BS. Without loss of generality, all users and the BS are assumed to be equipped with one single antenna. 

By performing channel coding and modulation, the transmitted symbol $x_k$ by user $k$ is spread onto $M$ orthogonal subcarriers by a unique spreading sequence $\boldsymbol{s_k} = (s_{1k}, s_{2k}, ..., s_{Mk})^T \in \mathbb{C}^M$. Particularly, we consider the overloaded NOMA system, i.e., $M<K$. For inactive users, their transmitted symbol is equal to zero. The $M$ signals received at the BS can be expressed individually as
\begin{equation}
    y_m = \sum_{k=1}^K g_{mk} s_{mk} x_k + n_m, \ \ m=1,2,..., M\ ,
\end{equation}
where $g_{mk}$ denotes the channel gain of user $k$ transmitted over the $m$th subcarrier and $n_m \sim \mathcal{CN}(0,\sigma^2)$ denotes the Gaussian noise with noise power $\sigma^2$. 
We adopt the Rayleigh fading channel model whose channel gains are independent and identically distributed (i.i.d.) complex Gaussian random variables, i.e., $g_{mk} \overset{i.i.d.}{\sim} \mathcal{CN}(0,1)\ \forall{m,k}$. To simplify the expression, we express the received signal vector $\boldsymbol{y} = (y_1, ..., y_M)^T$ as 

\vspace{-0.5em}
\begin{equation}\label{eq:y=Hx+v}
    \boldsymbol{y}=\boldsymbol{H}\boldsymbol{x}+\boldsymbol{n},
\end{equation}
where $\boldsymbol{x}=(x_1, ..., x_K)^T$ is the transmitted signal vector, $\boldsymbol{H}$ denotes the $M\times K$ channel matrix whose entries are $h_{mk} = g_{mk} s_{mk}$, and $\boldsymbol{n} = (n_1,...,n_M)^T$ is the Gaussian noise vector.

\subsection{Frame-wise Joint Sparsity Model}
Generally, users transmit data in consecutive time slots and remain active or inactive throughout the time frame~\cite{block_sparsity_MUD_NOMA_Du2018}~\cite{sparsity_blind_greedy_Yu2019}. Motivated by the temporal correlations in user activity, we extend the system from a single transmission to a time frame model, known as the \textit{frame-wise joint sparsity} model.

Given a time frame of length $J$, the common sparsity support $\mathcal{S}$ is defined as
\begin{equation}\label{eq:frame-wise support}
    supp(\boldsymbol{x}^{(1)}) = supp(\boldsymbol{x}^{(2)}) = \dots =supp(\boldsymbol{x}^{(J)}) \triangleq \mathcal{S},
\end{equation}
where $supp(\boldsymbol{x}^{(j)}) = \{k\ |\ x^{(j)}_k \neq 0, k\in\{1, ..., K\}\}$. The number of active users during each transmission is defined as $S = |\mathcal{S}|$. We thus achieve the \textit{frame-wise joint sparsity} model:
\begin{equation}\label{eq:joint sparse model Y=HX+N}
    \boldsymbol{Y}=\boldsymbol{H X+N},
\end{equation}
where $\boldsymbol{Y} = [\boldsymbol{y}^{(1)}, ..., \boldsymbol{y}^{(J)}] \in \mathbb{R}^{M\times J}$, $\boldsymbol{X} = [\boldsymbol{x}^{(1)}, ..., \boldsymbol{x}^{(J)}] \in \mathbb{R}^{K\times J}$ and $\boldsymbol{N} = [\boldsymbol{n}^{(1)}, ..., \boldsymbol{n}^{(J)}] \in \mathbb{R}^{M\times J}$. We restrict the length of the time frame to be shorter than the channel coherence time, so that the channel matrix $\boldsymbol{H} \in \mathbb{R}^{M\times K}$ remains constant throughout the entire time frame.

The MUD problem now becomes a 2-dimensional CS problem, where the estimation target is the signal matrix $\boldsymbol{X}$ with known channel matrix $\boldsymbol{H}$ and the received signal matrix $\boldsymbol{y}$, which could be formulated as solving the optimization problem:
\begin{equation}\label{eq:x=argmin(y-Hx)}
    \underset{\boldsymbol{X}}{\text{min}} ||\boldsymbol{Y}-\boldsymbol{H} \boldsymbol{X}||_2^2.
\end{equation}

Existing recovery methods consist of two major approaches, $\ell_1$-minimization and greedy algorithms. 
$\ell_1$-minimization methods~\cite{BasisPursuit_1994} provide high recovery accuracy and theoretical performance guarantees but suffer from heavy computational complexity and are sensitive to noise. Greedy algorithms~\cite{OMP_Donoho__2012} have relatively lower complexity but usually require a large number of measurements for exact recovery.


\vspace{-0.5em}
\subsection{Performance Metrics}
Symbol error rate (SER) has been used as the main evaluation method for MUD algorithms. It describes the general detection ability of the algorithm and fails to reflect some specific targets of MUD. To achieve a more thorough performance evaluation, we adopt three separate performance metrics: SER, detection probability ($P_d$) and false alarm probability ($P_{fa}$). Given the recovered symbol $\hat{X}^{(j)}_k$ for user $k$ in the $j$th time slot, the performance metrics are introduced as follows:
\begin{itemize}
    \item SER: SER is defined as the ratio of incorrectly recovered symbols transmitted by the active users to all symbols transmitted by the active users, given by 
    \vspace{-0.1em}
    \begin{equation}
        \text{SER}=\frac{1}{S} \sum_{k\in \mathcal{S}} \mathds{1}_{\{X^{(j)}_k \neq \hat{X}^{(j)}_k\}}.
    \end{equation}
    \item $P_d$: 
    $P_d$ is defined as the ratio of correctly detected active users to all active users, given by
    \vspace{-0.1em}
    \begin{equation}
        P_d = \frac{1}{S} \sum_{k\in \mathcal{S}} \mathds{1}_{\{\hat{X}^{(j)}_k \neq 0\}}.
    \end{equation}
    \item $P_{fa}$: $P_{fa}$ is defined as the ratio of incorrectly detected inactive users to all inactive users, given by
    \begin{equation}
        P_{fa} = \frac{1}{K-S} \sum_{k\notin \mathcal{S}} \mathds{1}_{\{\hat{X}^{(j)}_k \neq 0\}}.
    \end{equation}
\end{itemize}.



\section{Generative Networks for Multi-user Detection}\label{sec:generative network for MUD}
In this section, we first describe the MAML training procedure of generative networks based on the algorithm developed in~\cite{DCS_DeepMind}, then introduce the proposed network architecture, followed by the proposed GenMUD framework.

\vspace{-0.2em}
\subsection{Generative Network Training Algorithm}
\vspace{-0.2em}
MUD aims to obtain the user activity and decode the transmitted signal if the user is active, which could be achieved by solving~\eqref{eq:x=argmin(y-Hx)}. To achieve higher accuracy and lower complexity, we propose to use generative networks for solving~\eqref{eq:x=argmin(y-Hx)}, which is described in the following.

Instead of directly solving~\eqref{eq:x=argmin(y-Hx)}, a map is built from a latent space to the space of all possible transmitted signals as
\begin{equation}{\label{eq:x=G(z)}}
    \boldsymbol{X} = G_{\theta} (\boldsymbol{z}),
\end{equation}
where $\boldsymbol{z}$ represents a point in the latent space of arbitrary dimensions and $G_{\theta}$ is a generative neural network. By searching for a particular $\boldsymbol{\hat{z}}$ in the latent space that maps to an accurate estimation $\boldsymbol{\hat{X}}$, the MUD problem is reformulated as:
\begin{equation}{\label{eq:z=argmin(y-HG(z))}}
    \boldsymbol{\hat{z}} = \underset{\boldsymbol{z}}{\text{argmin}} ||\boldsymbol{Y}-\boldsymbol{H} G_{\theta}(\boldsymbol{z})
    ||_2^2\ .
\end{equation}
One common solution to the above non-convex optimization problem is to apply gradient descent starting from a randomly sampled point $\boldsymbol{z} \sim \mathcal{N}(0,1)$. Therefore, we can update $\boldsymbol{\hat{z}}$ by
\begin{equation}\label{eq:gradient descent of z}
    \boldsymbol{\hat{z}} \leftarrow \boldsymbol{\hat{z}} - \alpha \frac{\partial ||\boldsymbol{Y}-\boldsymbol{H}G_{\theta}(\boldsymbol{z})
    ||_2^2}{\partial \boldsymbol{z}},
\end{equation}
where $\alpha$ indicates the learning rate. However, it usually requires hundreds or even thousands of gradient descent steps until~\eqref{eq:gradient descent of z} converges. Fortunately, a separate loss function for $G_{\theta}$ has been proposed and minimized in conjunction with~\eqref{eq:x=G(z)}~\cite{DCS_DeepMind}. 
This technique of optimizing an optimization procedure is known as Model-Agnostic Meta-Learning (MAML)~\cite{MAML_1987}.

To formulate MAML in the context of MUD, we denote $p_{\text{task}}(\mathcal{T})$ as the distribution of tasks where each task $\mathcal{T}_i$ is to find an optimal $\boldsymbol{\hat{z}}_i$ that approximates the target signal matrix $\boldsymbol{X}_i$. MAML is employed by training the network weights $\boldsymbol{\theta}$ against the measurement error over all tasks:
\begin{equation}{\label{eq:MAML loss}}
\underset{\theta}{\text{min}}\ \mathcal{L}_G, \ \ \text{for}\ \mathcal{L}_G=\mathbb{E}_{\mathcal{T}_i \sim p_{\text{task}} (\mathcal{T})}\left[||\boldsymbol{Y}_i - \boldsymbol{H}_i G_{\theta}(\boldsymbol{\hat{z}}_i)
||_2^2\right],
\end{equation}
where $\boldsymbol{\hat{z}}_i$ is the output of task $\mathcal{T}_i$, i.e., the gradient descent steps in~\eqref{eq:gradient descent of z}. 
To further increase the convergence speed, the same optimization procedure is applied to the learning rate $\alpha$.

An additional constraint is needed for $G_{\theta}$ since it can exploit~\eqref{eq:MAML loss} and easily reach small loss by mapping all $G_{\theta}(\boldsymbol{\hat{z}})$ into the null space of $\boldsymbol{H}$. To address this problem, the generator is trained against Restricted Isometry Property (RIP) through an approximated RIP loss: 
\begin{equation}\label{eq: RIP loss}
    \mathcal{L}_H = \mathbb{E}_{\boldsymbol{X}_1, \boldsymbol{X}_2} \left[\Big(\big|\big|\boldsymbol{H} \boldsymbol{X}_1 - \boldsymbol{H} \boldsymbol{X}_2 \big|\big|_2 - \big|\big|\boldsymbol{X}_1-\boldsymbol{X}_2\big|\big|_2\Big)^2\right],
\end{equation}
where $\boldsymbol{X}_1$ and $\boldsymbol{X}_2$ are sampled from different stages in the training process. Specifically, we sample one true signal, two generated signals each before and after the operation in~\eqref{eq:gradient descent of z}, then compute an average loss over the three pairs of signals. Algorithm~\ref{Algorithm: train} illustrates the training procedure for $G_{\theta}$.


\begin{algorithm}
\caption{Generative Network Training Algorithm}
\begin{algorithmic}[1]
\label{Algorithm: train}
    \renewcommand{\algorithmicrequire}{\textbf{Input:}}
    \renewcommand{\algorithmicensure}{\textbf{Output:}}
    \REQUIRE Training data $\{\boldsymbol{X}_i\}_{i=1}^{N_d}$, channel matrix $\{\boldsymbol{H}_i\}_{i=1}^{N_d}$, generator $G_{\theta}$, number of latent update steps $T$
    \ENSURE  Trained generator $G_{\hat{\theta}}$, optimized learning rate $\hat{\alpha}$ 
 \\ Initialize $\theta$, $\alpha$
    \REPEAT
        \FOR {$i=1$ to $N_d$}
            \STATE Measure the signal $\boldsymbol{y}_i \leftarrow \boldsymbol{H}_i\boldsymbol{x}_i + \boldsymbol{n}_i$
            \STATE Sample $\boldsymbol{\hat{z}}_{i} \sim \mathcal{N}(0, \boldsymbol{I})$
            \FOR {$t = 1$ to $T$}
                \STATE $\boldsymbol{\hat{z}}_i \leftarrow \boldsymbol{\hat{z}}_i -  \alpha \frac{\partial}{\partial \boldsymbol{\hat{z}}}_i||\boldsymbol{y}_i-\boldsymbol{H}G_{\theta}(\boldsymbol{\hat{z}}_i)
                ||_2^2$
            \ENDFOR
        
        \STATE $\mathcal{L}_G = \frac{1}{N_d}\sum_{i=1}^{N_d} ||\boldsymbol{y}_i-\boldsymbol{H}G_{\theta}(\boldsymbol{\hat{z}}_i)
        ||_2^2$
        \STATE Compute $\mathcal{L}_H$ \text{using Eq.~\eqref{eq: RIP loss}}
        \STATE Update $\theta \leftarrow \theta - \frac{\partial}{\partial \theta} (\mathcal{L}_G + \mathcal{L}_H)$
        \STATE Update $\alpha \leftarrow \alpha - \frac{\partial}{\partial \alpha} (\mathcal{L}_G + \mathcal{L}_H)$
        \ENDFOR
    \UNTIL{reaches the maximum training steps}
    \STATE \textbf{Return} $G_{\hat{\theta}}$, $\hat{\alpha}$
\end{algorithmic} 
\end{algorithm}

\vspace{-1em}
\subsection{Proposed Neural Network Architecture}
Standard generative neural networks~\cite{DCGAN_Radford2016} are mainly designed for image data to extract spatial correlations and hidden structures. To be applied to MUD data which consists of random data points, larger input dimensions are required for greater learning capability. However, the use of fully-connected layers causes network sizes to increase dramatically with input dimensions, leading to high computational complexity. Moreover, the large kernel sizes used in typical networks introduce redundant network parameters when applied to MUD data, and have the risk of overfitting. Hence, we propose a low-complexity generative network designed specifically for MUD problems.

\begin{figure}
    \centering
    \includegraphics[width=3in]{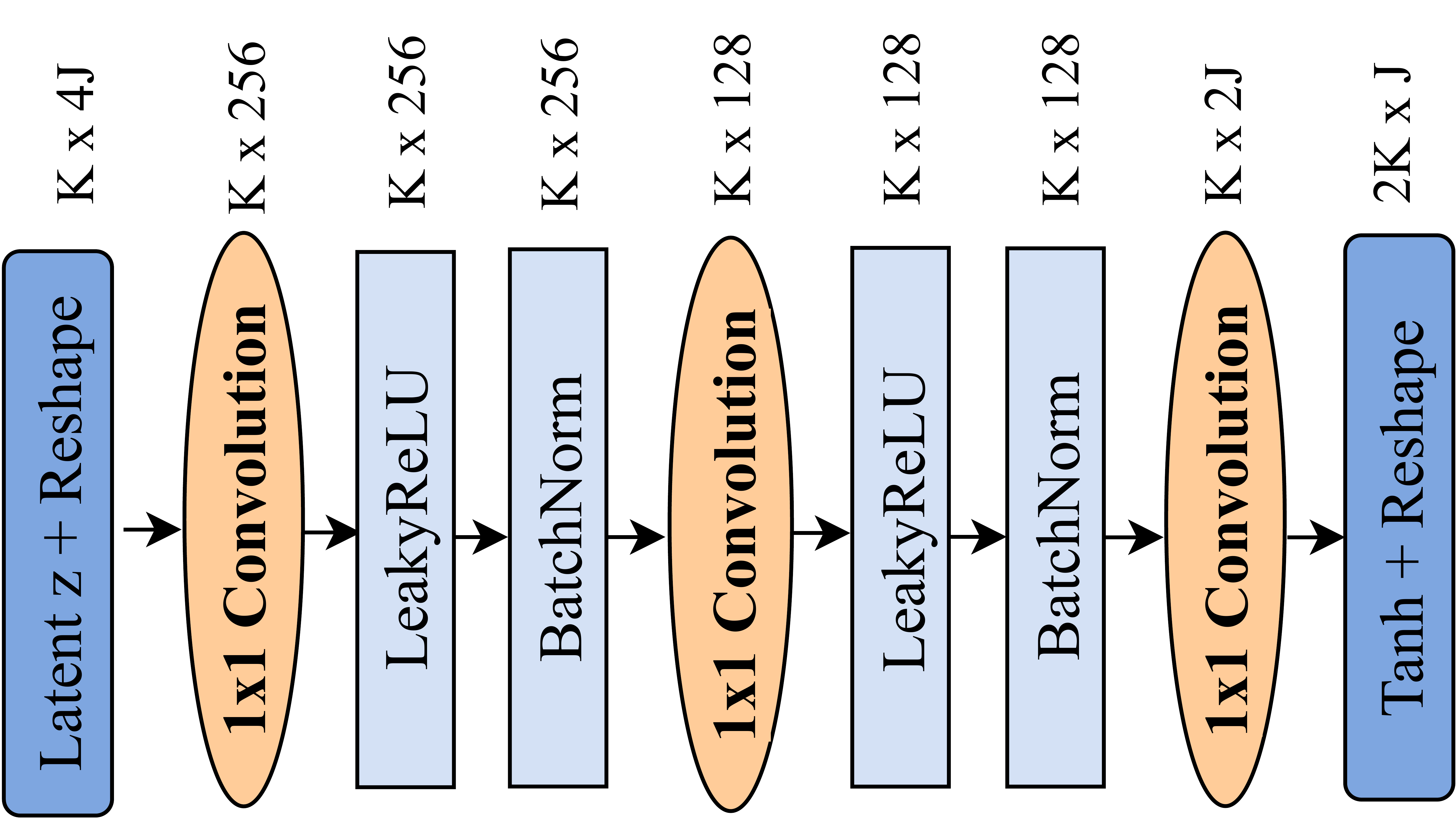}
    \caption{Structure of the adopted neural network.}
    \label{fig:network_sturcture}
\end{figure}

The proposed network has an input dimension of $4KJ \times 1$ and an output dimension of $2K\times J$, i.e., $[\Re(\boldsymbol{X}), \Im(\boldsymbol{X})]^T$. The hidden layers consist of three 1D convolutional layers all using a kernel size of 1. LeakyReLU activation and batch-normalization are employed after each convolutional layer except the last. The last layer employs Tanh activation to scale the outputs to [-1, 1], as displayed in Fig~\ref{fig:network_sturcture}.

The proposed architecture has three main differences compared to standard generative networks:
\vspace{-0.2em}
\begin{itemize}
    \item \textbf{Large input dimensions}: For generative neural networks, larger input dimensions are proved to increase the learning capability of the network at the cost of computational complexity~\cite{size_noise_tradeoff_GAN_Bailey2018}. 
    Given the considered model, we choose a latent dimension of twice the size of the output, since a further increase in the dimension improves accuracy by a negligible amount. 
    \item \textbf{No fully-connected layers}: Fully-connected layers scale badly with input sizes, hence are too expensive for solving MUD problems in term of complexity. In the proposed network, the input layer is followed directly with a convolutional layer to increase the model scalability as the input dimensions grow.
    \item \textbf{1x1 convolutional layers}: Considering the independent user behaviours, we propose to eliminate any spatial convolutions among the users. In particular, we replace all the 2D convolutional layers with 1x1 1D convolutional layers of input size $K$, i.e., the total number of users. The resulting network is capable of generating the framewise joint sparsity feature without any inter-user correlation.
\end{itemize}

\subsection{GenMUD Framework}
Given the received signals $\boldsymbol{Y}$, channel response matrix $\boldsymbol{H}$ and sparsity level $S$, the MUD problem can be solved by the trained model, including the trained generator $G_{\hat{\theta}}$ and the optimized learning rate $\hat{\alpha}$.

A set of initial reconstructions are obtained by optimizing the latent representation $\boldsymbol{z}$ from a random start using~\eqref{eq:gradient descent of z}. Both the dimension of $\boldsymbol{z}$ and the number of gradient descent steps $T$ must be the same values used in the training algorithm. The generator $G_{\hat{\theta}}$ is treated as fixed during the latent updates and $\hat{\alpha}$ is used as the learning rate.

Since neural networks output continuous numbers, we further map the outputs into valid modulation symbols. We calculate the magnitude of each reconstructed signal and, in each time slot, map the $S$ signals with greater magnitudes to the nearest constellation symbol, and the rest to zero indicating inactive users. Algorithm~\ref{Algorithm: test} illustrates the proposed GenMUD algorithm.

\begin{algorithm}
   \caption{GenMUD Algorithm}
   \begin{algorithmic}[1]
   \label{Algorithm: test}
   \renewcommand{\algorithmicrequire}{\textbf{Input:}}
    \renewcommand{\algorithmicensure}{\textbf{Output:}}
    \REQUIRE Received signals $\boldsymbol{Y}$, channel matrix $\boldsymbol{H}$, pre-trained generator $G_{\hat{\theta}}$, number of latent update steps $T$, optimized learning rate $\hat{\alpha}$
    \ENSURE  Reconstructed signals $\boldsymbol{\hat{X}}$
    \\ \STATE Sample $\boldsymbol{\hat{z}} \sim \mathcal{N}(0, \boldsymbol{I})$
        \FOR {$t = 1$ to $T$}
            \STATE $\boldsymbol{\hat{z}} \leftarrow \boldsymbol{\hat{z}} -  \hat{\alpha}\frac{\partial}{\partial \boldsymbol{\hat{z}}}||\boldsymbol{Y}-\boldsymbol{H}G_{\hat{\theta}}(\boldsymbol{\hat{z}})
            ||_2^2$
        \ENDFOR
        \STATE Initial reconstructions $\boldsymbol{\Tilde{X}} = G_{\hat{\theta}} (\boldsymbol{\hat{z}})$
        \STATE Initialize output $\boldsymbol{\hat{X}} = \boldsymbol{0}_{K\times J}$
        \FOR {$j = 1$ to $J$}
        \FOR {$s = 1$ to $S$}
            \STATE $p = \text{argmax}||\boldsymbol{\Tilde{X}}^{(j)}||^2_2$
            \STATE $\hat{X}^{(j)}_p = \{\Tilde{X}^{(j)}_p  \text{mapped to the nearest symbol}\}$
            \STATE $\Tilde{X}^{(j)}_p = 0$
        \ENDFOR
        \ENDFOR
 \STATE \textbf{Return} $\boldsymbol{\hat{X}}$
 \end{algorithmic} 
 \end{algorithm}

\vspace{-0.2em}
\section{Simulations}\label{sec:experiment}
In this section, we evaluate the performance of the proposed GenMUD algorithm in solving MUD problems compared to several existing solutions including orthogonal matching pursuit (OMP)~\cite{OMP_Donoho__2012}, a dynamic CS-based MUD method (DyCS)~\cite{DyCS_Bichai_Wang_2016} and basis pursuit de-noising (BPDN) algorithm~\cite{BasisPursuit_1994}. The oracle least squares (LS) algorithm, which assumes full knowledge of the sparse support, including the sparsity level and the location of sparse support, is considered as a benchmark method. For all considered methods, user sparsity is assumed to be known at the BS unless otherwise stated. In the simulations, the total number of users is $K=200$ and the transmitted signals are modulated by Quadrature Phase Shift Keying (QPSK). The number of consecutive time slots is fixed at $J=7$. Toeplitz random matrix is used to design the spreading matrix so that RIP is satisfied with high probability~\cite{Toeplitz_2009}.

\subsection{Model Training}
The proposed neural network is trained by the simulated dataset where the number of active users is $S=40$ among the $K=200$ users, the length of the time frame is $J=7$, SNR is set to 20 dB and the number of subcarriers is $M=100$. Latent parameters are normalized before the first update and the total number of latent update steps is $T=20$
. The initial value for the latent learning rate $\alpha$ is 0.01. Adam is used as the optimizer with learning rate 0.0001. The batch size is 32.

\subsection{Simulation Results for the MUD}
\begin{figure}
    \centering
    \includegraphics[width=3.3in]{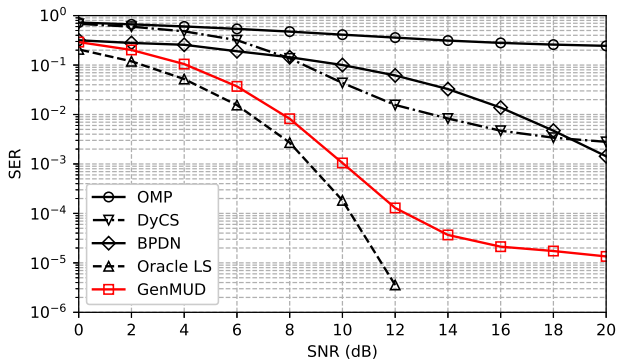}
    \caption{SER performance comparison against SNR.}
    \label{fig:SER_SNR}
\end{figure}

Fig.~\ref{fig:SER_SNR} depicts the SER performance comparison of the considered algorithms under different SNRs. In the simulations, the number of active users is $S=40$ and the number of subcarriers is $M=100$
. In Fig.~\ref{fig:SER_SNR}, the proposed GenMUD algorithm outperforms OMP and DyCS significantly in terms of SER over the whole range of SNR. In comparison to the BPDN approach, GenMUD shows increasing performance gain as SNR increases from 0 dB to 14 dB. It is observed that the proposed GenMUD demonstrates near oracle performance when SNR is lower than 8 dB and is the only method other than LS that achieves an SER lower than $1\times 10^{-3}$. Given that SNR is 20 dB during training, the results indicate that the neural network has learned to minimize the influence of noise during detection and provide high symbol detection accuracy through iterative latent updates.

\begin{figure}
    \centering
    \includegraphics[width=3.3in]{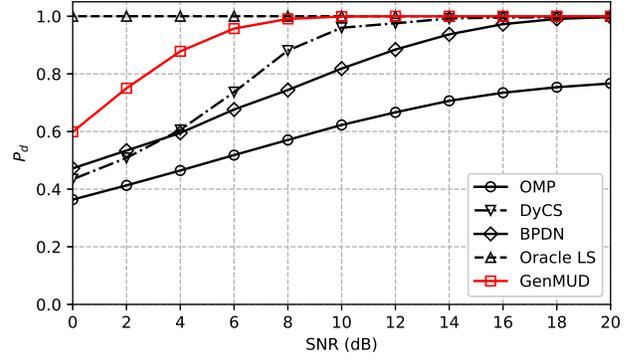}
    \setlength{\abovecaptionskip}{-1pt}
    \caption{Detection probability $P_d$ performance comparison against SNR.}
    \label{fig:Pd_SNR}
\end{figure}

\begin{figure}
    \centering
    \includegraphics[width=3.3in]{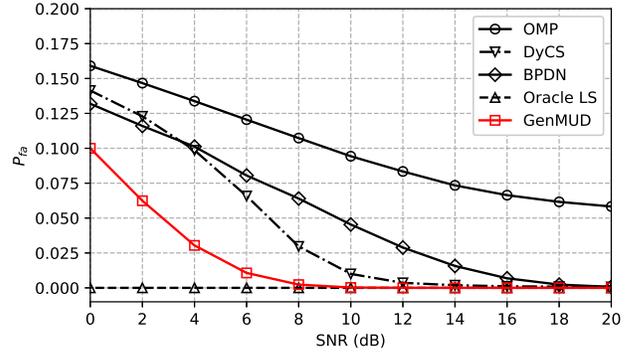}
    \setlength{\abovecaptionskip}{-1pt}
    \caption{False alarm probability $P_{fa}$ performance comparison against SNR.}
    \label{fig:Pfa_SNR}
\end{figure}

Fig.~\ref{fig:Pd_SNR} illustrates the detection performance, $P_d$, against SNRs for different methods. The number of active users is $S=40$ and the number of subcarriers is $M=100$. From the figure, the proposed GenMUD method achieves the highest $P_d$ among all methods, except for oracle LS which assumes full knowledge of the sparse support. As SNR becomes above 10 dB, the proposed GenMUD achieves nearly 100\% detection probability.

Fig.~\ref{fig:Pfa_SNR} compares the performance of the false alarm probability, $P_{fa}$, against SNRs among the considered methods. It is noted that the proposed GenMUD achieves the lowest $P_{fa}$ compared to OMP, BPDN and DyCS, and approaches zero $P_{fa}$ as SNR increases beyond 10 dB. The results in Fig.~\ref{fig:Pd_SNR} and Fig.~\ref{fig:Pfa_SNR} validate that the proposed GenMUD is capable of accurately identifying active users from inactive users for all SNRs. 

\begin{figure}
    \centering
    \includegraphics[width=3.3in]{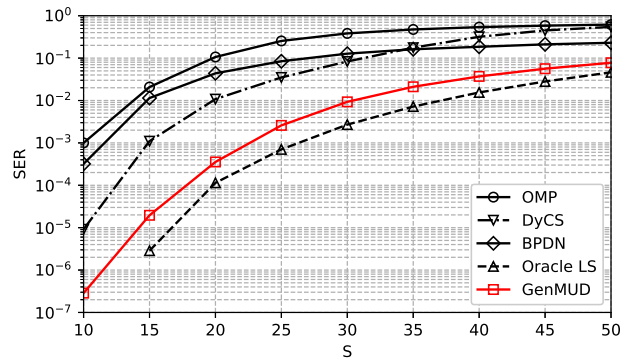}
    \setlength{\abovecaptionskip}{-2pt}
    \caption{SER performance comparison against user sparsity.}
    \label{fig:SER_sparsity}
\end{figure}

Fig,~\ref{fig:SER_sparsity} illustrates the effects of user sparsity on SER performance of the considered methods under $\text{SNR}=6$ dB and $M=100$. For all the methods, the SER performance degrades as the number of active users increases. However, the proposed GenMUD exhibits consistently lower SER than OMP, DyCS and BPDN throughout the range of SNR. Given that the proposed neural network is trained under $S=40$ active users, the consistent performance gains of the proposed method 
imply that the network has precisely captured the underlying relationships between user activity and the received signals.

\vspace{-1.5em}
\subsection{Sparsity Estimation}
\vspace{-0.5em}
Sparsity estimation has been a challenging topic in MUD due to the difficulty in measuring user sparsity prior to MUD
. Most MUD algorithms, such as DyCS and OMP, are designed based on known user sparsity which limits their applications in practice. Recently, some sparsity-blind MUD algorithms are proposed with different sparsity-blind strategies including approximation algorithms for sparsity built inside the MUD algorithms~\cite{block_sparsity_MUD_NOMA_Du2018} or stopping criteria for greedy algorithms with no prior knowledge of user sparsity~\cite{sparsity_blind_greedy_Yu2019}. These sparsity approximation methods are computationally expensive and do not apply to other MUD frameworks.


Based on extensive experiment results and data analysis, we propose a sparsity estimator given by
\begin{equation}
    \hat{S} = \mathbb{E}\left[\frac{\tau}{2(\tau+1)} ||\boldsymbol{y}||_2^2\right],
\end{equation}
where $\tau$ is SNR in its linear scale and $\boldsymbol{y}$ is the $M\times 1$ signal vector received in a single time slot. In terms of a framewise model of length $J$, the estimator is formulated as
\vspace{-0.2em}
\begin{equation}\label{eq:Khat_framewise}
    \hat{S} = \frac{1}{J}\sum_{j=1}^J \frac{\tau}{2(\tau+1)} ||\boldsymbol{y}^{(j)}||_2^2.
\end{equation}
The estimation accuracy is evaluated by the normalized error ($E_n$), given by
\vspace{-0.3em}
\begin{equation}
    E_n = \frac{|S - \hat{S}|}{S}.
\end{equation}

\begin{figure}
    \centering
    \includegraphics[width=3.4in]{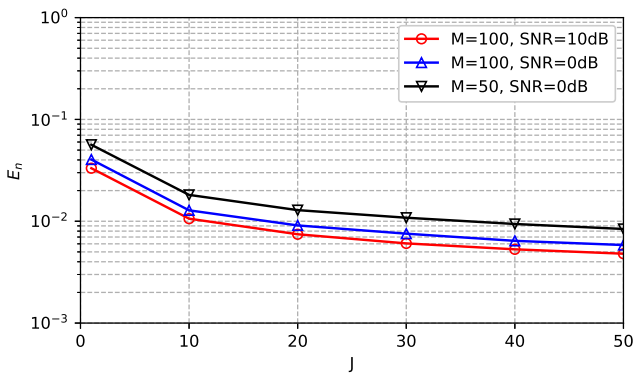}
    \setlength{\abovecaptionskip}{-1pt}
    \caption{Normalized error ($E_n$) of the sparsity estimator against frame length.}
    \label{fig:Khat_J}
\end{figure}

\begin{figure}
    \centering
    \includegraphics[width=3.3in]{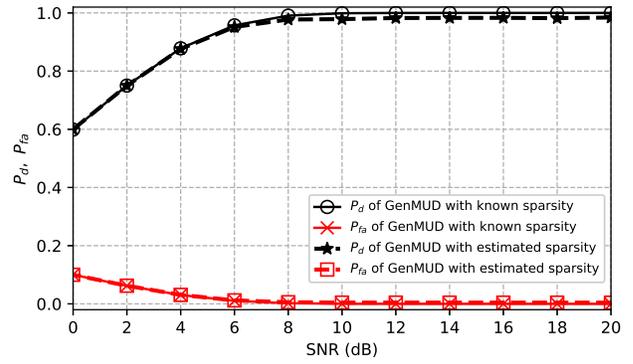}
    \setlength{\belowcaptionskip}{-8pt}
    \setlength{\abovecaptionskip}{-0.5pt}
    \caption{Detection probability ($P_d$) and false alarm probability ($P_{fa}$) performance against SNR of the proposed GenMUD algorithm with known sparsity or estimated sparsity.}
    \label{fig:Khat_Pd_Pfa}
\end{figure}

Fig.~\ref{fig:Khat_J} depicts the $E_n$ performance against frame length $J$ of the proposed sparsity estimator in the framewise model. Results are plotted for a different number of subcarriers $M$ and SNRs. As the frame length increases from 1 to 50, $E_n$ of the proposed estimator decreases gradually from around 0.05 to below 0.01 indicating a generally low estimation error. The decrease in $E_n$ also indicates an inverse relationship between estimation error and the number of time slots $J$. In practical scenarios, the BS can perform a constant update of the estimated sparsity as new signals are received to reduce error overtime. It is also observed that there is little increase in $E_n$ as the communication environment becomes severe, i.e., as $M$ decreases from 100 to 50 and SNR decreases from 10 dB to 0 dB, which demonstrates the robustness of the proposed estimator to the varying environment. %

Fig.~\ref{fig:Khat_Pd_Pfa} illustrates the $P_d$ and $P_{fa}$ performance of the proposed GenMUD with known sparsity and estimated sparsity. 
In the simulations, the number of active users is $S = 40$ and the number of subcarriers is $M = 100$. It is observed that, the proposed GenMUD with estimated sparsity achieves the same $P_{fa}$ performance as when sparsity is known. For SNR smaller than 6 dB, the sparsity estimator is used with no impact on $P_d$ performance, and for SNR greater than 6 dB, the performance difference is unnoticeable.
Fig.~\ref{fig:Khat_Pd_Pfa} shows that the proposed sparsity estimator provides accurate approximation and has a neglectable influence on support detection performance.

\section{Conclusions}\label{sec:conclusion}
In this paper, we investigated the problem of multi-user detection (MUD) for grant-free non-orthogonal multiple access (NOMA) using compressed sensing (CS) techniques. We proposed a generative neural network-based MUD (GenMUD) framework and designed a generative neural network to achieve higher learning capability with lower complexity by exploiting the independent user behaviours. Simulation results showed that the proposed GenMUD method provided better detection performance compared to conventional CS and MUD approaches in terms of SER, detection probability and false alarm probability. We also provided a sparsity estimator 
supported with simulation results demonstrating low estimation error and negligible impact on support detection accuracy under various communication settings.

\bibliographystyle{IEEEtran}
\bibliography{Yixuan_ICC_submitted}

\end{document}